\documentclass[aps,pra,twocolumn,showpacs]{revtex4}
\usepackage{amssymb}

\usepackage{bm}

\begin{document}

\title{Entanglement conditions for tripartite systems via indeterminacy relations}
\author{Lijun Song$^{1,2,3}$, Xiaoguang Wang$^{1}$, Dong Yan$^{2,4}$, and
Zhong-Sheng Pu$^{4}$,}\affiliation{1,Zhejiang Institute of Modern
Physics, Department of Physics, Zhejiang University, HangZhou
310027, P.R.China.}\affiliation{2,Institute of
Applied Physics, Changchun University, Changchun 130022, P.R.China } %
\affiliation{3, School of Science, Changchun University of Science
and Technology, Changchun 130022, P.R.China} \affiliation{4,School
of Science, Lanzhou University of Technology, Lanzhou 730050,
P.R.China} \ 
\date{\today}

\begin{abstract}
Based on the Schr{\"{o}}dinger-Robertson indeterminacy relations in
conjugation with the partial transposition, we derive a class of
inequalities for detecting entanglement in several tripartite
systems, including bosonic, SU(2), and SU(1,1) systems. These
inequalities are in general stronger than those based on the usual
Heisenberg relations for detecting entanglement. We also discuss the
reduction from SU(2) and SU(1,1) to bosonic systems and the
generalization to multipartite case.
\end{abstract}

\maketitle

\narrowtext

\section{Introduction}

The Heisenberg uncertainty relation (HUR) plays a fundamental role
in quantum mechanics, and recent developments in quantum information
theory display that it is useful for deriving some entanglement
criteria ~\cite
{Shchukin1,Agarwal,Nha1,Nha2}. Given two noncommuting observables \{$A$, $B$%
\} satisfying $[A,B]=C$, the HUR is given by \cite{Heisenberg}
\begin{equation}
\langle (\Delta A)^2\rangle \langle (\Delta B)^2\rangle \ge \frac 14|\langle
C\rangle |^2,
\end{equation}
where Var$(A)\equiv \langle (\Delta A)^2\rangle =\langle A^2\rangle -\langle
A\rangle ^2$ denotes the variance or the uncertainty of the observable $A.$
It is evident that the product of two uncertainties is bounded below by $%
|\langle C\rangle |^2/4.$

Actually, there exists a stronger bound $|\langle C\rangle |^2/4+$Cov$%
(A,B)^2,$ where the covariance Cov$(A,B)=$ $\langle \left( AB+BA\right)
/2\rangle -\langle A\rangle \langle B\rangle .$ The corresponding
uncertainty relation is the Schr{\"{o}}dinger-Robertson indeterminacy
relation (SRIR) given by \cite{SR1,SR2}
\begin{equation}
\langle (\Delta A)^2\rangle \langle (\Delta B)^2\rangle \ge \frac 14|\langle
C\rangle |^2+\text{Cov}(A,B)^2.
\end{equation}
Very recently, the SRIR was also used by Nha \textit{et al}.
\cite{Nha3} and Yu \textit{et al}. \cite{SixiaYu} to obtain
entanglement conditions. In general, the entanglement criteria based
on SRIRs are stronger than those via HURs.

Many methods are developed to obtain entanglement conditions in the
literature~\cite
{Hofmann,Toth,Raymer,Hillery,Mancini,Simon,Peres,Miranowicz,Horodecki,Giedke,Giovannetti,Klyachko}%
. The method based the uncertainty relations has its own advantages
for deriving entanglement criteria. It can apply to not only
continuous-variable but also discrete-variable systems, or even
hybrid systems. Another advantage is that it is easier to use to
derive entanglement criteria comparing with several other
approaches. Finally and importantly, the entanglement criteria based
on this method often provide strong detection of the separability.
For instance, the entanglement inequality based on SRIR for two
qubits gives a necessary and sufficient condition for separability~
\cite{SixiaYu}.

In this paper, we consider tripartite states and study their
separability problem via indeterminacy relations. Some separability
inequalities have been obtained previously in Ref. \cite{ZongGuoLi}
from a different approach. It will be seen that the inequalities
obtained here are more general and stronger. We consider not only
continuous-variable systems but also SU(2) and SU(1,1) systems.

\section{Method based on indeterminacy relations}

First, we introduce the method and demonstrate its usefulness by
re-deriving the inequality given by Duan \textit{et
al}.~\cite{DuanLM}. Consider the SRIRs for operators $A,B,C$ acting
on a composite multipartite system. The SRIR of course holds for a
separable state represented by the density operator $\rho $. The
separable state is still separable after partial transposition with
respect to any subsystems, namely the partially transposed density
operator $\rho ^{\text{pt}}$ is still physical. Thus, the SRIR also
holds for state
\begin{equation}
\langle (\Delta A)^2\rangle _{\rho ^{\text{pt}}}\langle (\Delta B)^2\rangle
_{\rho ^{\text{pt}}}\ge \frac 14|\langle C\rangle _{\rho ^{\text{pt}}}|^2+%
\text{Cov}(A,B)_{\rho ^{\text{pt}}}^2.
\end{equation}
This is of the form of product of two uncertainties. By using
$a^2+b^2\geq 2ab,$ one can also achieve the following
\begin{eqnarray}
&&\alpha \langle (\Delta A)^2\rangle _{\rho ^{\text{pt}}}+\beta \langle
(\Delta B)^2\rangle _{\rho ^{\text{pt}}}  \nonumber \\
&\ge &\sqrt{\alpha \beta }\sqrt{|\langle C\rangle _{\rho ^{\text{pt}}}|^2+4%
\text{Cov}(A,B)_{\rho ^{\text{pt}}}^2},
\end{eqnarray}
which is of the form of arbitrary sum of two uncertainties. Here,
$\alpha ,\beta $ are real. By defining positive $c=\sqrt{\beta
/\alpha },$ the above equation can be written as
\begin{eqnarray}
&&\langle (\Delta A)^2\rangle _{\rho ^{\text{pt}}}+c^2\langle (\Delta
B)^2\rangle _{\rho ^{\text{pt}}}  \nonumber \\
&\ge &c\sqrt{|\langle C\rangle _{\rho ^{\text{pt}}}|^2+4\text{Cov}%
(A,B)_{\rho ^{\text{pt}}}^2}.
\end{eqnarray}

For any operators $A,$\ acting on a state $\rho $, we have\textit{\ }
\begin{equation}
\langle A\rangle _{\rho ^{\text{pt}}}=\langle A^{\text{pt}}\rangle _\rho .
\end{equation}
Then, using this fact, inequalities (3) and (4) can be written in the form
of partial transposition of operators other than states. They are given by
\begin{eqnarray}
&&\left[ \langle \left( A^2\right) ^{\text{pt}}\rangle _\rho -\langle A^{%
\text{pt}}\rangle _\rho ^2\right] \times \left[ \langle \left( B^2\right) ^{%
\text{pt}}\rangle _\rho -\langle B^{\text{pt}}\rangle _\rho ^2\right]
\nonumber \\
&\ge &\frac 14|\langle C^{\text{pt}}\rangle _\rho |^2+\left[ \langle
[A,B]_{+}^{\text{pt}}/2\rangle _\rho -\langle A^{\text{pt}}\rangle _\rho
\langle B^{\text{pt}}\rangle _\rho \right] ^2,  \nonumber \\
&&  \label{ine1}
\end{eqnarray}
and
\begin{eqnarray}
&&\left[ \langle \left( A^2\right) ^{\text{pt}}\rangle _\rho -\langle A^{%
\text{pt}}\rangle _\rho ^2\right] +c^2\left[ \langle \left( B^2\right) ^{%
\text{pt}}\rangle _\rho -\langle B^{\text{pt}}\rangle _\rho ^2\right]
\nonumber \\
&\ge &c\sqrt{|\langle C^{\text{pt}}\rangle _\rho |^2+4\left[ \langle
[A,B]_{+}^{\text{pt}}/2\rangle _\rho -\langle A^{\text{pt}}\rangle _\rho
\langle B^{\text{pt}}\rangle _\rho \right] ^2},  \nonumber \\
&&  \label{ine2}
\end{eqnarray}
respectively. Here, $[A,B]_{+}=AB+BA.$ Note that in general $\left(
A^2\right) ^{\text{pt}}\neq \left( A^{\text{pt}}\right) ^2,(AB)^{\text{pt}%
}\neq B^{\text{pt}}A^{\text{pt}}.$ The inequalities hold for any separable
states, and conversely any state violating this inequality must be entangled.

Now, we re-derive the inequality for two-mode system given by Duan
\textit{et al}~\cite{DuanLM} using the present approach. Consider
the following operators
\begin{equation}
u^{\prime }=|a|x_1+\frac 1ax_2,v^{\prime }=|a|p_1+\frac 1ap_2,
\end{equation}
where $x_i$ and $p_i$ are position and momentum operators for mode $i$,
respectively. It is easy to check that for any state they two operators
satisfy the HUR
\begin{equation}
\langle (\Delta u^{\prime })^2\rangle \langle (\Delta v^{\prime })^2\rangle
\ge \frac 14(a^2+\frac 1{a^2})^2,
\end{equation}
Therefore, we have
\begin{equation}
\langle (\Delta u^{\prime })^2\rangle +\langle (\Delta v^{\prime })^2\rangle
\ge a^2+\frac 1{a^2},
\end{equation}
holding for any state. For a separable state $\rho $, we have $\langle
(\Delta u^{\prime })^2\rangle _{\rho ^{\text{T}_2}}+\langle (\Delta
v^{\prime })^2\rangle _{\rho ^{\text{T}_2}}\ge a^2+\frac 1{a^2},$ where T$_2$
denotes the partial transposition with respect to the second mode. By noting
the fact that $p^{\text{T}}=-p,x^{\text{T}}=x,$ we then obtain

\begin{equation}
\langle (\Delta u)^2\rangle +\langle (\Delta v)^2\rangle \ge a^2+\frac
1{a^2}.
\end{equation}
for any separable states. Here, $u=u^{\prime },v=|a|p_1-\frac
1ap_2.$ We see that from the uncertainty relation in conjugation
with the partial transposition, the inequality by Duan et al. is
neatly obtained, indicating the effectiveness of the approach.

\section{Entanglement conditions for tripartite systems}

We consider entanglement of tripartite systems, and begin our
discussions on the case of three bosonic modes.

\subsection{Continuous-variable systems}

Let operators $a$, $b$, and $c$ be the annihilation operators of the first
(A), second (B), and third (C) mode. We define a set of operators $L_x,$ $%
L_y $ and $L_z$ which obey the commutation relations
$[L_x,L_y]=iL_z$. Note that these three operators do not need to
form an algebra. It can be realized in optics using three-mode
fields represented by the annihilation operators,
\begin{eqnarray}
L_x &=&\frac 12(a^{\dagger }b^{\dagger }c+abc^{\dagger }), \nonumber\\
L_y &=&\frac 1{2i}(a^{\dagger }b^{\dagger }c-abc^{\dagger }), \nonumber\\
L_z &=&\frac 12\left[ N_aN_b(N_c+1)-(N_a+1)(N_b+1)N_c\right] ,
\end{eqnarray}
where $N_a=a^{\dagger }a$, $N_b=b^{\dagger }b$, and $N_c=c^{\dagger }c$. We
further define another set of operators $H_x,$ $H_y$ and $H_z$ that satisfy $%
[H_x,H_y]=iH_z$. The operators can be given by
\begin{eqnarray}
H_x &=&\frac 12(a^{\dagger }b^{\dagger }c^{\dagger }+abc), \nonumber\\
H_y &=&\frac 1{2i}(a^{\dagger }b^{\dagger }c^{\dagger }-abc), \nonumber\\
H_z &=&\frac 12\left[ N_aN_bN_c-(N_a+1)(N_b+1)(N_c+1)\right] .
\end{eqnarray}

It is easy to see that the two sets of operators are connected by partial
transposition with respect to the third mode as follows
\begin{eqnarray}
H_x^{\text{T}_3} &=&L_x,H_y^{\text{T}_3}=L_y,  \nonumber \\
H_z^{\text{T}_3} &=&H_z,L_z^{\text{T}_3}=L_z.  \label{pt1}
\end{eqnarray}
The partial transposition with respect to the third mode means that
we are considering the entanglement between systems $AB$ and $C.$
From the discussions in the above section, in order to get
entanglement conditions, we need to know the partial transposition
of product of two operators. For our case, after some algebras, we
obtain
\begin{eqnarray}
\left( H_x^2\right) ^{\text{T}_3} &=&L_x^2+\frac 14(N_a+N_b+1),  \nonumber \\
\left( H_y^2\right) ^{\text{T}_3} &=&L_y^2+\frac 14(N_a+N_b+1),  \nonumber \\
\left( \lbrack H_x,H_y]_{+}\right) ^{\text{T}_3} &=&[L_x,L_y]_{+}.
\label{pt2}
\end{eqnarray}

Now by replacing $A,$ $B,$ and $C$ in Eq. (\ref{ine1}) with $H_x,$ $H_y$,
and $H_z,$ respectively, and using Eqs. (\ref{pt1}) and (\ref{pt2}), we
obtain the following inequality
\begin{eqnarray}
&&\left[ \langle \Delta L_x\rangle _\rho ^2+\frac 14\langle N_a+N_b+1\rangle
\right]  \nonumber \\
&&\times \left[ \langle \Delta L_y\rangle _\rho ^2+\frac 14\langle
N_a+N_b+1\rangle \right]  \nonumber \\
&\geqslant &\frac 14|\langle H_z\rangle |^2+\text{Cov}(L_x,L_y)^2.  \nonumber
\\
&=&\frac 1{16}\left[ \langle M_{+}\rangle +\langle N_{+}\rangle +1\right] ^2+%
\text{Cov}(L_x,L_y)^2.  \label{hw}
\end{eqnarray}
where $M_{+}=N_aN_b+N_bN_c+N_aN_c,$ $N_{+}=N_a+N_b+N_c.$ Violation of the
inequality gives a sufficient condition for $AB|C$ entanglement.

To connect our results with inequalities previously obtained in the
literature, we apply Eq. (\ref{ine2}) to the present three mode
case, and then obtain
\begin{eqnarray}
&&\langle \Delta L_x\rangle _\rho ^2+\langle \Delta L_y\rangle _\rho ^2+%
\frac{\left( 1+c^2\right) }4\langle N_a+N_b+1\rangle  \nonumber \\
&\geqslant &c\sqrt{\frac 14\left[ \langle M_{+}\rangle +\langle N_{+}\rangle
+1\right] ^2+4\text{Cov}(L_x,L_y)^2.}
\end{eqnarray}
For $c=1,$ the above equation reduces to
\begin{eqnarray}
&&\langle \Delta L_x\rangle _\rho ^2+\langle \Delta L_y\rangle _\rho ^2
\nonumber \\
&\geqslant &\sqrt{\frac 14\left[ \langle M_{+}\rangle +\langle N_{+}\rangle
+1\right] ^2+4\text{Cov}(L_x,L_y)^2.}  \nonumber \\
&&-\frac 12\langle N_a+N_b+1\rangle .  \label{c1}
\end{eqnarray}
If we use HUR other than SRIR, one has
\begin{equation}
\langle \Delta L_x\rangle _\rho ^2+\langle \Delta L_y\rangle _\rho
^2\geqslant \frac 12\langle M_{+}+N_c\rangle .  \label{c2}
\end{equation}
by letting Cov$(L_x,L_y)=0.$ This inequality is just the one
obtained from a different procedure \cite{ZongGuoLi}. Inequality
(\ref{c2}) is a special case of inequality (\ref{c1}). Having
studied three-mode systems, we next consider the SU(2) spin systems
and SU(1,1) systems.

\subsection{SU(2) spin and SU(1,1) systems}

\subsubsection{SU(2) spin systems}

A spin is described by the operators $J_{\pm }$ and $J_z,$ which obeys the
following commutation relations
\begin{equation}
\left[ J_{+},J_{-}\right] =2J_z,[J_z,J_{\pm }]=\pm J_{\pm }.
\end{equation}
In the spin system, we can define the `number' operator $\mathcal{N}=J_z+j.$
For tripartite systems, we define
\begin{eqnarray}
A_x &=&\frac 12(J_{a+}J_{b+}J_{c-}+J_{a-}J_{b-}J_{c+}),  \nonumber \\
A_y &=&\frac 1{2i}(J_{a+}J_{b+}J_{c-}-J_{a-}J_{b-}J_{c+}),  \nonumber \\
A_z &=&\frac 12[J_{a+}J_{a-}J_{b+}J_{b-}J_{c-}J_{c+}  \nonumber \\
&&-J_{a-}J_{a+}J_{b-}J_{b+}J_{c+}J_{c-}],
\end{eqnarray}
satisfying $[A_x,A_y]=iA_z.$ By using
\begin{eqnarray}
J_{+}J_{-} &=&\mathcal{N}(2j-\mathcal{N}+1),  \nonumber \\
J_{-}J_{+} &=&(\mathcal{N}+1)(2j-\mathcal{N}),  \label{pm}
\end{eqnarray}
operator $A_z$ can be written as
\begin{eqnarray}
A_z &=&\frac 12[\mathcal{N}_a\mathcal{N}_b(\mathcal{N}_c+1)(2j_a-\mathcal{N}%
_a+1)  \nonumber \\
&&\times (2j_b-\mathcal{N}_b+1)(2j_c-\mathcal{N}_c)  \nonumber \\
&&-(\mathcal{N}_a+1)(\mathcal{N}_b+1)\mathcal{N}_c(2j_a-\mathcal{N}_a)
\nonumber \\
&&\times (2j_b-\mathcal{N}_b)(2j_c-\mathcal{N}_c+1)].
\end{eqnarray}
Another set of operators satisfying $\left[ B_x,B_y\right] =iB_z$ are given
by
\begin{eqnarray}
B_x &=&\frac 12(J_{a+}J_{b+}J_{c+}+J_{a-}J_{b-}J_{c-}),  \nonumber \\
B_y &=&\frac 1{2i}(J_{a+}J_{b+}J_{c+}-J_{a-}J_{b-}J_{c-}),  \nonumber \\
B_z &=&\frac 12[J_{a+}J_{a-}J_{b+}J_{b-}J_{c+}J_{c-}  \nonumber \\
&&-J_{a-}J_{a+}J_{b-}J_{b+}J_{c-}J_{c+}]
\end{eqnarray}
By using Eq. (\ref{pm}), operator $B_z$ can be written as
\begin{eqnarray}
B_z &=&\frac 12[\mathcal{N}_a\mathcal{N}_b\mathcal{N}_c(2j_a-\mathcal{N}_a+1)
\nonumber \\
&&\times (2j_b-\mathcal{N}_b+1)(2j_c-\mathcal{N}_c+1)  \nonumber \\
&&-(\mathcal{N}_a+1)(\mathcal{N}_b+1)(\mathcal{N}_b+1)  \nonumber \\
&&\times (2j_a-\mathcal{N}_a)(2j_b-\mathcal{N}_b)(2j_c-\mathcal{N}_c)].
\end{eqnarray}

From the definitions of above operators, one finds
\begin{eqnarray}
B_x^{\text{T}_3} &=&A_x,B_y^{\text{T}_3}=A_y,  \nonumber \\
B_z^{\text{T}_3} &=&B_z,A_z^{\text{T}_3}=A_z,  \nonumber \\
\left( B_x^2\right) ^{\text{T}_3} &=&A_x^2+\frac 14E,  \nonumber \\
\left( B_y^2\right) ^{\text{T}_3} &=&A_y^2+\frac 14E,  \nonumber \\
\left( \lbrack B_x,B_y]_{+}\right) ^{\text{T}_3} &=&[A_x,A_y]_{+},
\end{eqnarray}
where
\begin{eqnarray}
E &=&2(\mathcal{\mathcal{N}}_c-j_c)[\mathcal{N}_a\mathcal{N}_b(2j_a-\mathcal{%
N}_a+1)(2j_b-\mathcal{N}_b+1)  \nonumber \\
&&-(\mathcal{N}_a+1)(\mathcal{N}_b+1)(2j_a-\mathcal{N}_a)(2j_b-\mathcal{N}%
_b)].  \label{eee}
\end{eqnarray}
Then, from Eq. (\ref{ine1}), we obtain
\begin{eqnarray}
&&\left[ \langle \Delta A_x\rangle ^2+\frac 14\langle E\rangle \right]
\left[ \langle \Delta A_y\rangle ^2+\frac 14\langle E\rangle \right]
\nonumber \\
&\ge &\frac 14|\langle B_z\rangle |^2+\text{Cov}(A_x,A_y)^2.  \label{su2}
\end{eqnarray}
This is the entanglement condition for tripartite SU(2) systems and
can be used to detect entanglement between $AB$ and $C$.

\subsubsection{SU(1,1) systems}

The SU(1,1) systems are described by the su(1,1) Lie algebra. The generators
of su(1,1) Lie algebra, $K_z$ and $K_{\pm },$ satisfy the commutation
relations
\begin{equation}
\lbrack K_{+},K_{-}]=-2K_z,\text{ }[K_z,K_{\pm }]=\pm K_{\pm }.
\end{equation}
Its discrete representation is
\begin{eqnarray}
K_{+}|m,k\rangle &=&\sqrt{(m+1)(2k+m)}|m+1,k\rangle ,  \nonumber \\
K_{-}|m,k\rangle &=&\sqrt{m(2k+m-1)}|m-1,k\rangle ,  \nonumber \\
K_z|m,k\rangle &=&(m+k)|m,k\rangle .  \label{rep}
\end{eqnarray}
Here $|m,k\rangle (m=0,1,2,...)$ is the complete orthonormal basis and $%
k=1/2,1,3/2,2,...$ is the Bargmann index labeling the irreducible
representation [$k(k-1)$ is the value of Casimir operator]. We introduce the
`number' operator $\mathcal{M}$ by
\begin{equation}
\mathcal{M}=K_z-k,\mathcal{M}|m,k\rangle =m|m,k\rangle .
\end{equation}
From Eq. (\ref{rep}), one may find
\begin{eqnarray}
K_{+}K_{-} &=&\mathcal{M}(2k+\mathcal{M}-1),  \nonumber \\
K_{-}K_{+} &=&(\mathcal{M}+1)(2k+\mathcal{M}).  \label{pm2}
\end{eqnarray}

Similar to the discussions of SU(2) case, we consider the $AB|C$
entanglement conditions for three SU(1,1) systems. We define
\begin{eqnarray}
C_x &=&\frac 12(K_{a+}K_{b+}K_{c-}+K_{a-}K_{b-}K_{c+}),  \nonumber \\
C_y &=&\frac 1{2i}(K_{a+}K_{b+}K_{c-}-K_{a-}K_{b-}K_{c+}),  \nonumber \\
C_z &=&\frac 12[K_{a+}K_{a-}K_{b+}K_{b-}K_{c-}K_{c+}  \nonumber \\
&&-K_{a-}K_{a+}K_{b-}K_{b+}K_{c+}K_{c-}],
\end{eqnarray}
satisfying $[C_x,C_y]=iC_z.$ By using Eq. (\ref{pm2}), operator $C_z$ can be
written as
\begin{eqnarray}
C_z &=&\frac 12[\mathcal{M}_a\mathcal{M}_b(\mathcal{M}_c+1)(2k_a+\mathcal{M}%
_a-1)  \nonumber \\
&&\times (2k_b+\mathcal{M}_b-1)(2k_c+\mathcal{M}_c)  \nonumber \\
&&-(\mathcal{M}_a+1)(\mathcal{M}_b+1)\mathcal{M}_c(2k_a+\mathcal{M}_a)
\nonumber \\
&&\times (2k_b+\mathcal{M}_b)(2k_c+\mathcal{M}_c-1)].
\end{eqnarray}
Another set of operators satisfying $\left[ D_x,D_y\right] =iD_z$ are given
by
\begin{eqnarray}
D_x &=&\frac 12(K_{a+}K_{b+}K_{c+}+K_{a-}K_{b-}K_{c-}),  \nonumber \\
D_y &=&\frac 1{2i}(K_{a+}K_{b+}K_{c+}-K_{a-}K_{b-}K_{c-}),  \nonumber \\
D_z &=&\frac 12[K_{a+}K_{a-}K_{b+}K_{b-}K_{c+}K_{c-}  \nonumber \\
&&-K_{a-}K_{a+}K_{b-}K_{b+}K_{c-}K_{c+}]
\end{eqnarray}
Operator $D_z$ can be written in the form
\begin{eqnarray}
D_z &=&\frac 12[\mathcal{M}_a\mathcal{M}_b\mathcal{M}_c(2k_a+\mathcal{M}_a-1)
\nonumber \\
&&\times (2k_b+\mathcal{M}_b-1)(2k_c+\mathcal{M}_c-1)  \nonumber \\
&&-(\mathcal{M}_a+1)(\mathcal{M}_b+1)(\mathcal{M}_c+1)  \nonumber \\
&&\times (2k_a+\mathcal{M}_a)(2k_b+\mathcal{M}_b)(2k_c+\mathcal{M}_c)].
\end{eqnarray}

From the definitions of above operators, one finds

\begin{eqnarray}
D_x^{\text{T}_3} &=&C_x,D_y^{\text{T}_3}=C_y,  \nonumber \\
D_z^{\text{T}_3} &=&D_z,C_z^{\text{T}_3}=C_z,  \nonumber \\
\left( D_x^2\right) ^{\text{T}_3} &=&C_x^2+\frac 14F,  \nonumber \\
\left( D_y^2\right) ^{\text{T}_3} &=&C_y^2+\frac 14F,  \nonumber \\
\left( \lbrack D_x,D_y]_{+}\right) ^{\text{T}_3} &=&[C_x,C_y]_{+},
\end{eqnarray}
where
\begin{eqnarray}
F &=&2(\mathcal{\mathcal{M}}_c+k_c)[(\mathcal{M}_a+1)(\mathcal{M}_b+1)
\nonumber \\
&&\times (2k_a+\mathcal{M}_a)(2k_b+\mathcal{M}_b)  \nonumber \\
&&-\mathcal{M}_a\mathcal{M}_b(2k_a+\mathcal{M}_a-1)(2k_b+\mathcal{M}_b-1)].
\end{eqnarray}
Then, from Eq. (\ref{ine1}), we obtain
\begin{eqnarray}
&&\left[ \langle \Delta C_x\rangle ^2+\frac 14\langle F\rangle \right]
\left[ \langle \Delta C_y\rangle ^2+\frac 14\langle F\rangle \right]
\nonumber \\
&\ge &\frac 14|\langle D_z\rangle |^2+\text{Cov}(C_x,C_y)^2.  \label{su11}
\end{eqnarray}
It is known that su(2) and su(1,1) algebras connects with
Heisenberg-Weyl algebra, and thus we expect that the inequalities
for SU(2) and SU(1,1) systems also relates to the corresponding
inequality for bosonic systems.

\subsection{Reduction from SU(2) and SU(1,1) to bosons}

We use the usual Holstein-Primakoff realization of su(2) algebra \cite
{Holstein} :
\[
J_{+}=a^{\dagger }\sqrt{2j-a^{\dagger }a},\text{ }J_{-}=\sqrt{2j-a^{\dagger
}a}a,\text{ }J_z=a^{\dagger }a-j.
\]
In the limit of $j\rightarrow \infty ,$ we have
\[
\frac{J_{+}}{\sqrt{2j}}\rightarrow a^{\dagger },\text{ }\frac{J_{-}}{\sqrt{2j%
}}\rightarrow a,\text{ }-\frac{J_z}j\rightarrow 1.
\]
by expanding the square root and neglecting terms of $O(1/j).$
Holstein-Primakoff transformation \cite{Holstein} representation for the
su(1,1) algebra is given by

\[
K_{+}=a^{\dagger }\sqrt{2k+a^{\dagger }a},K_{-}=\sqrt{2k+a^{\dagger }a}a,%
\text{ }K_z=a^{\dagger }a+k.
\]
In the limit of $k\rightarrow \infty ,$ we have
\[
\frac{K_{+}}{\sqrt{2k}}\rightarrow a^{\dagger },\text{ }\frac{K_{-}}{\sqrt{2k%
}}\rightarrow a,\text{ }\frac{K_z}k\rightarrow 1,
\]
by expanding the square root and neglecting terms of $O(1/k).$ We see that
both the su(2) and su(1,1) algebras reduce to Heisenberg-Weyl algebra in the
large $j$ or $k$ limit.

Multiplying (\ref{su2}) with $1/(2j_12j_22j_3),$ and letting $%
j_1,j_2,j_3\rightarrow \infty ,$ we can see that
\begin{eqnarray}
\langle \Delta A_x\rangle ^2&\rightarrow &\langle \Delta L_x\rangle ^2, \nonumber\\
\langle \Delta A_y\rangle ^2 &\rightarrow &\langle \Delta L_y\rangle ^2, \nonumber\\
\text{Cov}(A_x,A_y) &\rightarrow &\text{Cov}(L_x,L_y), \nonumber\\
\langle B_z\rangle &\rightarrow &\langle H_z\rangle .
\end{eqnarray}
From Eq. (\ref{eee}), in this limit, we find that operator
$E\rightarrow N_a+N_b+1.$ Thus, inequality (\ref{su2}) for SU(2)
system reduces to inequality (\ref{hw}) for the bosonic system.
Similarity, in the limit of $k_1,k_2,k_3\rightarrow \infty ,$
inequality (\ref{su11}) for SU(1,1) system reduces to inequality
(\ref{hw}).

\section{Generalization to multipartite systems}

The methods employed above for tripartite states can be extended to $n$%
-partite states. For the sake of illustration, we consider $n$ modes
whose annihilation operators are give by $a_1,$ $a_2,\cdots $ and
$a_n,$ respectively, and study the entanglement between $n$-th mode
and the rest. We have two set of operators ,
\begin{eqnarray}
L_x &=&\frac 12(a_1^{\dagger }a_2^{\dagger }\cdots a_{n-1}^{\dagger
}a_n+a_1a_2\cdots a_{n-1}a_n^{\dagger }), \nonumber\\
L_y &=&\frac 1{2i}(a_1^{\dagger }a_2^{\dagger }\cdots a_{n-1}^{\dagger
}a_n-a_1a_2\cdots a_{n-1}a_n^{\dagger }), \nonumber\\
L_z &=&\frac 12\left[ (N_n+1)\stackrel{n-1}{\prod_{i=1}}N_i-N_n%
\prod_{i=1}^{n-1}(N_i+1)\right] ,
\end{eqnarray}
and
\begin{eqnarray}
H_x &=&\frac 12(a_1^{\dagger }a_2^{\dagger }\cdots a_n^{\dagger
}+a_1a_2\cdots a_n), \\
H_y &=&\frac 1{2i}(a_1^{\dagger }a_2^{\dagger }\cdots a_n^{\dagger
}-a_1a_2\cdots a_n), \\
H_z &=&\frac 12\left[ \stackrel{n}{\prod_{i=1}}N_i-\stackrel{n}{\prod_{i=1}}%
(N_i+1)\right] .
\end{eqnarray}
satisfying $[L_x,L_y]=iL_z,[H_x,H_y]=iH_z.$

From the definitions of above operators, one finds
\begin{eqnarray}
H_x^{\text{T}_n} &=&L_x,H_y^{\text{T}_n}=L_y,H_z^{\text{T}_n}=H_z,L_z^{\text{%
T}_n}=L_z,  \nonumber \\
\left( H_x^2\right) ^{\text{T}_n} &=&L_x^2+\frac 14\left( \stackrel{n-1}{%
\prod_{i=1}}(N_i+1)-\stackrel{n-1}{\prod_{i=1}}N_i\right) ,  \nonumber \\
\left( H_y^2\right) ^{\text{T}_n} &=&L_y^2+\frac 14\left( \stackrel{n-1}{%
\prod_{i=1}}(N_i+1)-\stackrel{n-1}{\prod_{i=1}}N_i\right) ,  \nonumber \\
\left( \lbrack H_x,H_y]_{+}\right) ^{\text{T}_n} &=&[L_x,L_y]_{+},
\end{eqnarray}
Then, from Eq. (\ref{ine1}), we obtain
\begin{eqnarray}
&&\left[ \langle \Delta L_x\rangle ^2+\frac 14\langle \stackrel{n-1}{%
\prod_{i=1}}(N_i+1)-\stackrel{n-1}{\prod_{i=1}}N_i\rangle \right]  \nonumber
\\
&&\times \left[ \langle \Delta L_y\rangle ^2+\frac 14\langle \stackrel{n-1}{%
\prod_{i=1}}(N_i+1)-\stackrel{n-1}{\prod_{i=1}}N_i\rangle \right]  \nonumber
\\
&\ge &\frac 1{16}\left[ \langle \stackrel{n}{\prod_{i=1}}(N_i+1)-\stackrel{n%
}{\prod_{i=1}}N_i\rangle \right] ^2+\text{Cov}(L_x,L_y)^2.  \nonumber \\
&&
\end{eqnarray}
This inequality is applicable to studies of entanglement properties between $%
n$-th mode and the rest. It is straightforward to obtain relevant
inequalities for entanglement between a finite selected modes and
the rest.

\section{Conclusions}

In summary, we have presented a family of entanglement criteria
which are able to detect entanglement in tripartite systems. The
method is based on the indeterminacy relations in conjugation with
the partial transposition. To detect entanglement, one need to
define appropriate two sets of operators, and write out the
indeterminacy relation in terms of the variances, covariances, and
expectation values. Then, after partial transposition on operators
other than states, we can obtain the entanglement criteria. One
merit of this method is that it is efficient to get useful strong
entanglement criteria.

We have considered three typical systems, bosonic, SU(2), and
SU(1,1) systems. We also discussed the reduction from SU(2) and
SU(1,1) to bosonic systems and the generalization to multipartite
case. We highlight the importance of uncertainty relations and the
indeterminacy relations. They are not only important in the
understanding of fundamental problems such as measurement problem in
quantum mechanics, but also provide a convenient way to detect
entanglement together with the partial transposition. We hope that
this work will stimulate more discussions on applications of the
indeterminacy relations in entanglement detection problems.

\acknowledgements X. Wang thanks for the valuable discussions with
Z. B. Chen, N. L. Liu, Z. W. Zhou, and C. P. Sun. This work is
supported by NSFC with grant Nos.10405019 and 90503003; NFRPC with
grant No. 2006CB921206; Specialized Research Fund for the Doctoral
Program of Higher Education (SRFDP) with grant No.20050335087.

\end{document}